\documentclass[12pt,preprint]{aastex}

\newcommand{\mdot}{M$_{\odot}$~}
\newcommand{\ldot}{L$_{\odot}$~}
\newcommand{\co}{\textrm{CO} }
\newcommand{\kmsec}{km~s$^{-1}$~}

\shorttitle{}
\shortauthors{Klamer et al.}

\begin{document}

\title{CO(1-0) \& CO(5-4) observations of the most distant known radio galaxy at $z=5.2$}


\author{I. J. Klamer$^{1,2}$}\email{i.klamer@physics.usyd.edu.au}
\author{R. D. Ekers$^{2}$}\email{rekers@csiro.au}
\author{E. M. Sadler$^{1}$}\email{e.sadler@physics.usyd.edu.au}
\author{A. Weiss$^{3}$}\email{aweiss@iram.es}
\author{R. W. Hunstead$^{1}$}\email{r.hunstead@physics.usyd.edu.au}
\author{C. De Breuck$^{4}$}\email{cdebreuc@eso.org}
\affil{1. School of Physics A28 University of Sydney, NSW 2006 Australia}%
\affil{2. CSIRO Australia Telescope National Facility, PO Box 76 Epping NSW 1710 Australia}%
\affil{3. IRAM, Avenida Divina Pastora 7, 18012 Granada, Spain}%
\affil{4. European Southern Observatory, Karl Schwarzchild Strasse 2, 85748 Garching bei Muenchen, Germany}%


\begin{abstract}
Using the Australia Telescope Compact Array we have detected \co(1-0) and \co(5-4) from TN~J0924--2201 at $z=5.2$, the most distant radio galaxy known to date. This is the second highest redshift detection of \co published so far. The \co(1-0) is $250-400$\kmsec wide with a peak flux density of $520\pm115~\mu$Jy~beam$^{-1}$ whilst the \co(5-4) line emission is $200-300$\kmsec wide with a peak flux density of $7.8\pm2.7~$mJy~beam$^{-1}$. Both transitions are spatially unresolved but there is marginal evidence for spatial offsets between the \co and the host galaxy; the \co(1-0) is located $28\pm11$~kpc ($4\farcs5\pm1\farcs7$) north of the radio galaxy whilst the \co(5-4) is located $18\pm8$~kpc ($2\farcs8\pm1\farcs2$) south of the radio galaxy. Higher spatial resolution observations are required to determine the reality of these offsets. Our result is the second detection of \co in a high redshift galaxy without pre-selection based on a massive dust content.   

\end{abstract}




\section{Introduction}
\label{intro}

Right out to the highest redshifts so far identified, radio galaxies are the most massive systems at each epoch. They reside in regions of over-density \citep{lef96,car97,pas98,pen00b}, corroborating evolutionary models which trace high redshift radio galaxies ($z>2$; HzRGs) through to nearby central dominant elliptical galaxies  in cluster-type environments (e.g. \citealp{lil84}). Unlike their nearby counterparts, however, HzRGs are not massive ellipticals with old stellar populations. Instead, their rest-frame optical/UV morphologies are composed of multiple blobs of clumpy emission often aligned along the axis of relativistic jets \citep{wvb98,pen99}. Rest frame far infra-red observations are strongly suggestive of large-scale, massive, dust-obscured star formation \citep{ste03}. Furthermore, HzRGS are often enshrouded by massive gas reservoirs which extend over tens of kpc, traced both by Lyman-$\alpha$ halos \citep{mcc90,vil03} and by carbon monoxide (\co) line emission \citep{pap00,cdb03b}. Each of these properties is consistent with the scenario where the galaxies are being observed in their formative stages. It is essential, therefore, that the characteristics of the HzRG population are investigated in detail in order to further our understanding of galaxy, and cluster, formation and evolution. \newline 
In this Letter we report the detection of molecular gas at $z=5.2$ in TN~J0924--2201, the highest redshift radio galaxy and the second highest redshift \co detection to date, after the $z=6.42$ quasar SDSS~J1148+5251 \citep{fan03,wal03}. Where necessary we use a flat $\Lambda$-dominated cosmology with $H_0=71$~\kmsec Mpc$^{-1}$ and $\Omega_{\Lambda}=0.73$ \citep{spe03}. In this cosmology, $1\arcsec=6.3$~kpc at $z=5.2$.

\section{TN~J0924--2201: selection \& characteristics}
 TN~J0924--2201 was initially discovered \citep{cdb00} from radio surveys using the well established technique for finding HzRGs of selecting ultra-steep spectrum sources with faint K-band counterparts (e.g. \citealp{cha96,blu98}). Follow-up optical spectroscopy detected faint Lyman-$\alpha$ emission from the host galaxy at a redshift of $5.1989\pm0.0006$ \citep{wvb99,ven04}. Based on its compact ($1\farcs3$) radio morphology, under-luminous Lyman-$\alpha$ emission compared with other HzRGs and rest frame 350~nm emission composed of several clumpy optical features, it was suggested that TN~J0924--2201 is a young, primeval galaxy actively forming the bulk of its stars \citep{wvb99}. 
We initially chose TN~J0924--2201 as a candidate for \co observations because, as the most distant known member of a sample of HzRGs, it is presumably one of the youngest \citep{blu99}. The existence of molecular gas would provide independent evidence for ongoing massive star formation, expected if we are probing its formation epoch. The southern declination ($\delta=-22^{\circ}$) of this source makes observations from millimetre telescopes in the northern hemisphere difficult, and its redshift moves all \co transitions out of any Very Large Array (VLA) band. We used the new wide bandwidth 1.2~cm and 3~mm receivers on the Australia Telescope Compact Array (ATCA) to search for \co(1-0) and \co(5-4) with $\nu_{\rm rest}=115.2712$~GHz and $576.2677$~GHz, respectively. \newline
\section{ATCA Observations}
\label{atcaobservations}
Observations were carried out with the ATCA during 2004 August and September. The ATCA is an array of six 22-m antennas with baselines ranging from 30~m to 6~km and capable of East-West, North-South and Hybrid array configurations. We utilised the maximum available bandwidth of the current correlator with 128~MHz/64 channel configurations for two adjacent IF bands. The spectral resolution of this configuration is 4.4~MHz. For the \co(1-0) observations, we used the hybrid H~168 array\footnote{The size of the naturally weighted restoring beam in the H~168 array at 18.6~GHz is $14\farcs5\times10\farcs1$, or $90\times63$~kpc.}, tuning the IFs to 18.561 and 18.631~GHz respectively, allowing an overlap region (after bandpass correction) of 20~MHz. This setup benefited from a simultaneous velocity coverage of $2500$~\kmsec and resolution of 72~\kmsec. System temperatures ranged from 30--50~K and phase stability was within $\pm20^{\circ}$ over the course of the observations. Phase, amplitude, bandpass and astrometric calibration were acquired by short observations of PKS~B0919--260 every 12--25 minutes depending on the weather. Absolute flux density calibration was ensured by snapshot observations of Mars. We estimate the calibration to be accurate to better than $\pm20\%$. 

The observational setup was repeated in the H~75 array\footnote{The size of the naturally weighted restoring beam in the H~75 array at 92.8~GHz is $5\farcs9\times4\farcs7$, or $37\times30$~kpc.} with the IFs tuned to 92.853 and 92.921~GHz respectively for the \co(5-4) transition. System temperatures (measured once per hour using a 300~K mechanical paddle moved over the feedhorn) ranged from 200--350~K and phase stability was mostly within $\pm50^{\circ}$ over the course of the observations. Phase and amplitude calibration were based on observations of PKS~B0834--201 every 12 minutes and the bandpass was determined by 10~minute observations of either PKS~B1253--055 or PKS~B0537--441. The rms error in the pointing ranged from $3-7\arcsec$ depending on the antenna (the primary beam at 93~GHz is $31\arcsec$). We periodically observed the 0.5~Jy quasar PKS~B0925--203, less than $2^{\circ}$ from TN~J0924--2201, to verify the calibration and astrometry. We estimate the flux scale to be accurate to better than $\pm30\%$. Data reduction was carried out in accordance with standard ATCA calibration procedures for millimetre  observing\footnote{http://www.atnf.csiro.au/computing/miriad}. Overall, $40$ hours  and $35$ hours of observing time were employed at 18.6~GHz and 92.9~GHz respectively to carry out this detection experiment. Table~\ref{tbl-1} gives a summary of our results.
\section{Results}

\subsection{\co(1-0)}
\begin{figure}
\centering
\includegraphics[width=8cm]{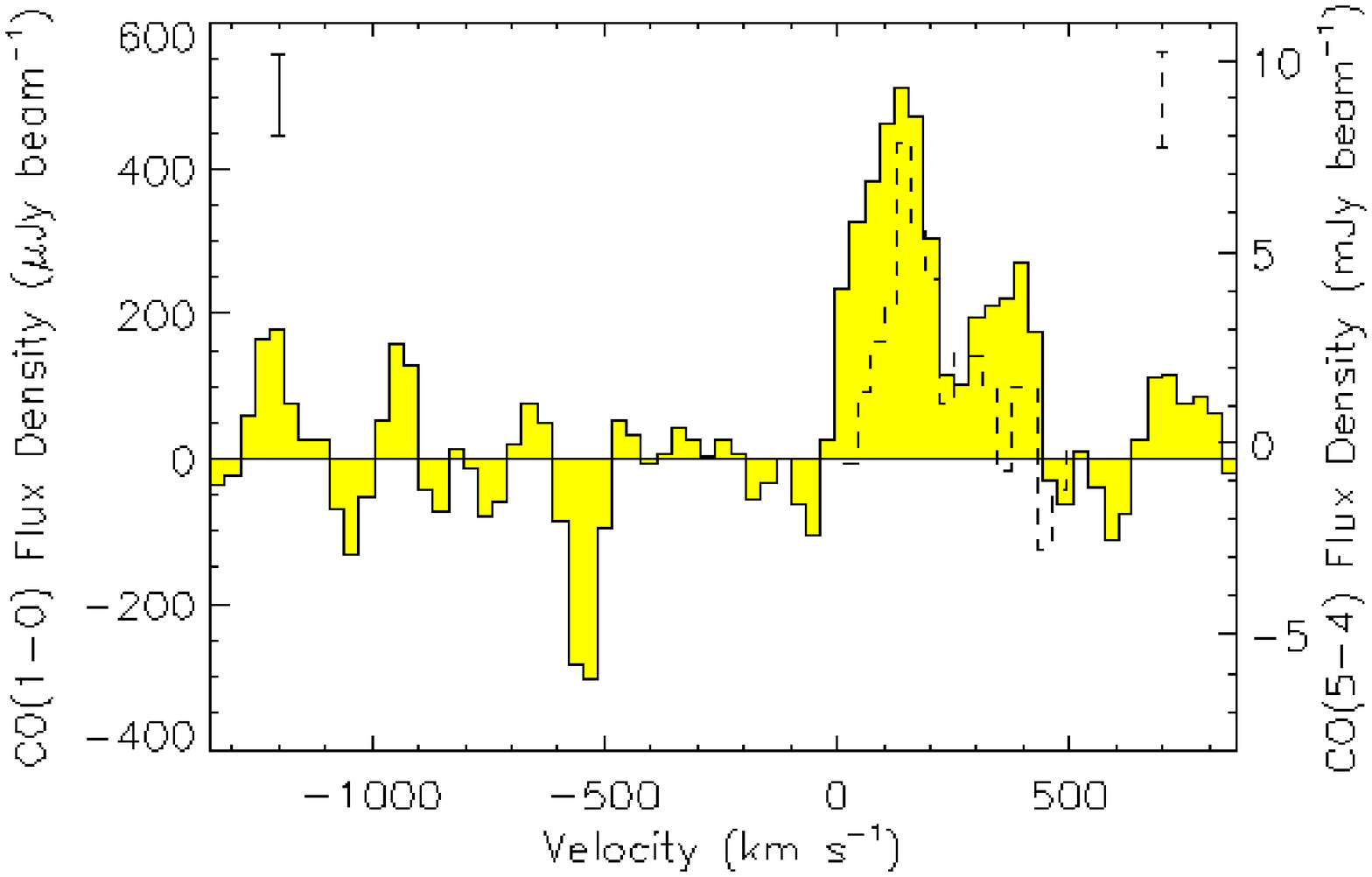}
\caption{\small Spectra of the continuum-subtracted \co(1-0) (shaded region) and \co(5-4) (dashed line) transitions, through the peak of their respective velocity-integrated emission shown in Figure~\ref{fig3}. The \co(1-0) spectrum is unbinned (32 \kmsec bins; spectral resolution 72~\kmsec) whilst the \co(5-4) spectral channels have been re-binned into 30 \kmsec bins (original spectral resolution 15~\kmsec). The velocity axis is defined with respect to the Lyman-$\alpha$ derived redshift of $z=5.1989$ and the bars in the top left and right corners show the rms noise level in the \co(1-0) and \co(5-4) cubes respectively.}
\label{fig1}
\end{figure}

\noindent The \co(1-0) line is detected in at least five independent channels covering $\sim400$~\kmsec and shown in Figure~\ref{fig1}. We subtracted a 0.71~mJy continuum in the UV plane using a first order spectral baseline fit. The continuum image is shown in greyscale in Figure~\ref{fig3}. The velocity-integrated, continuum-subtracted, flux density is $0.087\pm 0.017$~Jy~beam$^{-1}$~\kmsec and the corresponding intrinsic \co line luminosity is $L'_{\textsc{co}}=1.2\pm0.26\times10^{11}$~K~km~s$^{-1}$pc$^2$. There are two kinematic features apparent in the spectrum: the brighter, dominant, emission centred at $v=150$~\kmsec ($z=5.202\pm0.001$) and a weaker feature lying $\sim200$~\kmsec redward. Deeper observations are required to determine if the weaker feature is real. As commonly observed, the \co lies redward of the Lyman-$\alpha$ derived redshift (see Table 1 in \citealp{kla04}). There is no conclusive evidence for spatially extended gas, but the velocity-integrated emission (Figure~\ref{fig3}) appears to be offset to the north of the radio galaxy by $4\farcs5\pm1\farcs7$ ($28\pm11$~kpc). At this stage we do not consider this offset to be significant, since it is much less than a beam size away from the radio galaxy position. For comparison, the position angle of the projected radio axis is $29^{\circ}$ (see Figure~\ref{fig3}). Deeper observations with higher spatial resolution are required to better constrain the spatial extent of this gas.
 
\subsection{\co(5-4)}
\label{54}
At 93~GHz our instantaneous velocity coverage was only $600$~\kmsec with a spectral resolution of 15~\kmsec. The resultant spectrum is shown in Figure~\ref{fig1} and the emission profile integrated from 0 to 300 \kmsec is shown in Figure~\ref{fig3}. Although the zero level in the profile is well determined from offset positions in the image, our bandwidth is only marginally wider than the line width, so we have to consider the possibility of a spectral baseline offset due to continuum emission from the radio galaxy. However, the flux density we observe is more than 100 times that expected from the continuum. This is illustrated by the radio to sub-millimetre spectral energy distribution shown in Figure~\ref{fig4}. Extrapolating the non-thermal synchrotron spectrum with a spectral index of $\alpha=-1.7$, we expect a contribution of $<0.04$ mJy whilst extrapolating from the 352~GHz SCUBA limit assuming a modified blackbody spectrum with $\beta=1.5$ and T=50~K, the expected contribution due to thermal dust radiation is  $<0.01$ mJy. 
The most likely source of the 93~GHz emission is therefore the redshifted \co(5-4) transition at $z=5.202$. The velocity-integrated flux density is $1.19\pm0.27$~Jy~beam$^{-1}$~\kmsec, corresponding to $L'_{\textsc{co}}=4.3\pm1.0\times10^{10}$~K~km~s$^{-1}$~pc$^2$. Again, the emission appears offset from the radio galaxy position, the peak lying $2\farcs8\pm1\farcs2$ ($18\pm8$~kpc) south. Again, the positional offset is less than a beam size. Deeper observations are required to constrain the spatial extent of the molecular gas, including whether there are physically distinct gas emitting regions.

\section{Physical properties of the molecular gas}
\subsection{Mass Estimates}
\label{mass}

\co observations are translated into total H$_2$ mass estimates, assuming that the observed \co traces molecular clouds on large scales and that the molecular gas mass dominates the total dynamical mass of the galaxy. These mass estimates are uncertain at best, but to facilitate comparison with observations from other HzRGs, we assume M$(\textrm{H}_2)/L'_{\textsc{co}}\equiv\alpha=0.8$ \mdot~(K~\kmsec pc$^{2}$)$^{-1}$ as measured for nearby ultraluminous infra-red galaxies \citep{dow98}. Since molecular clouds provide both the fuel and the site for the next generation of star formation, our result implies that there is $\sim10^{11}$~\mdot of available gas from which to manufacture the next stellar generation. This estimate is consistent with M$(\textrm{H}_2)$ estimates of other HzRGs like 4C~60.07 \citep{gre04,pap00} and B3~J2330+3927 \citep{cdb03b}.  There is no direct evidence to indicate that TN~J0924--2201 is gravitationally lensed. However, if lensing is important then $L'_{\textsc{co}}$, and hence M$(\textrm{H}_2)$, are upper limits only.
\subsection{Temperature and Density from LVG modelling}
\label{mass}

The ratio of the peak flux densities would be equal to the square of the ratio of the upper transition levels (1/25) in the event that both lines are optically thick and thermally excited. Under the assumption of optically thick gas, the ratio we measure is 1/($15\pm9$), which suggests the \co(5-4) transition is sub-thermal; however, the errors are large. We have used a standard single-component large velocity gradient (LVG) code, with a typical CO-to-H$_2$ abundance X[CO]/(dv/dr)=$8\times10^{-5}$~(\kmsec)$^{-1}$~pc, to further constrain the physical properties of the molecular gas.  A good match to the observations is given by log(n(H$_2))=3.3$ (where n(H$_2$) is measured in units of cm$^{-3}$), T$_{\rm kin}=50$~K and r$>2.0$~kpc, but other density/temperature combinations are also consistent with the observed fluxes (see Figure~\ref{fig5}). Observations of \co transitions higher than \co(5-4) are required to constrain the excitation conditions of the gas. If the observed offsets between the \co(1-0) and \co(5-4) transitions reflect physically distinct \co emitting regions, then the excitation conditions presented here become invalid. 
\label{LVG}\subsection{Where is the dust?}
At present, the most efficient technique for detecting high redshift molecular gas is to pre-select the dustiest objects, as determined from their large rest frame far-infra red luminosities using (sub)millimetre sensitive instruments like SCUBA on the JCMT or MAMBO on the IRAM 30m telescope. This also means that current samples are biased towards objects with the largest dust-to-gas ratios ($L_{\textsc{fir}}/L'_{\textsc{co}}\sim300$ \ldot~K$^{-1}$~km$^{-1}$~s~pc$^{-2}$; \citealp{car04b}). For comparison, the median ratio is 160 for nearby ultra-luminous infra red galaxies with $L_{\textsc{fir}}\gtrsim10^{12}$ \ldot \citep{sol97} and about 50 for nearby luminous infra red galaxies with $10^{11} $\ldot$<L_{\textsc{fir}}\lesssim10^{12}$\ldot \citep{gao04}. The selection bias at high redshift is further compounded by the fact that initial \co detections are usually made in $J\geq3$ order lines, which prejudices the sample against regions of low density molecular gas where the higher order transitions are sub-thermally excited.\newline

Five of the six published \co detections of HzRGs constitute the most FIR-luminous $z>1$ radio galaxies recently surveyed with SCUBA \citep{reu04}. In this sample, approximately 50\% of the 24 observed galaxies were not detected; TN~J0924--2201 featured amongst these non-detections with $L_{850\mu m}<1\times10^{24}$~W~Hz$^{-1}$, five times underluminous compared to the \co detected HzRGs in the sample. Therefore, TN~J0924--2201 will have a dust-to-gas ratio at least five times smaller than the current sample of high redshift galaxies, similar to the nearby luminous infra-red galaxy population. The only other \co emitter at high redshift not selected on the basis of FIR emission --- that we are aware of --- is 53W002 at $z=2.394$ \citep{yam95,sco97,all00}. This source has a $3\sigma$ upper limit of 4.3~mJy at 850$\mu$m \citep{arc01}, corresponding to $L_{850\mu m}<7\times10^{24}$~W~Hz$^{-1}$. An a posteriori detection of the dust content in 53W002 ($1.7\pm0.4$ mJy at 1.3~mm; \citealp{all00}) suggests that a similar observation of TN~J0924--2201 may also reveal its dust content. Thus 53W002 and TN~J0924--2201 may be representative of a population of high redshift sources which remain undetected due to current selection techniques.
  
\section{The co-evolution of a central Dominant and a protocluster}
\begin{figure}
\centering
\rotatebox{270}{\includegraphics[width=6cm]{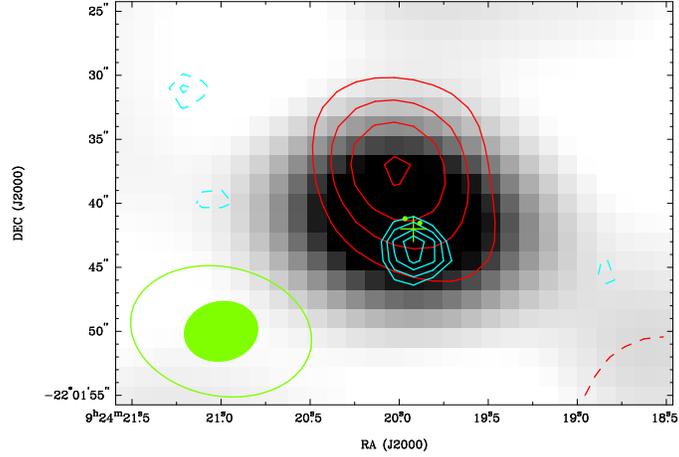}}
\caption{\small  Greyscale 18.6~GHz continuum image with \co(1-0) and \co(5-4) velocity-integrated emission overlaid in red and cyan respectively. The continuum flux density is $0.71\pm0.03$~mJy, the \co(1-0) contour levels are $\pm 2, 3, 4, 5 \times\sigma$  where $\sigma=0.017$~Jy~beam$^{-1}$\kmsec and the \co(5-4) contour levels are $\pm 2.5, 3, 3.5, 4 \times\sigma$ where $\sigma=0.27$~Jy~beam$^{-1}$\kmsec. The sizes of the 18~GHz (open ellipse) and 93~GHz (filled ellipse) restoring beams are shown in the bottom left corner, the cross shows the optical position of the radio galaxy. The radio lobes, illustrated with filled circles, extend $1\farcs3$ along a position angle of $29^{\circ}$.}
\label{fig3}
\end{figure}

\begin{figure}
\centering
\includegraphics[width=8cm]{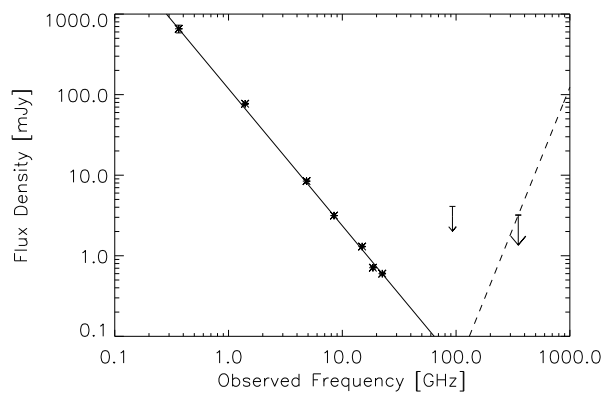}
\caption{\small Radio to sub-millimetre spectral energy distribution of TN~J0924--2201. The stars are radio observations from the Texas, NVSS and PMN radio sky-surveys along with pointed VLA observations and the ATCA 18.6~GHz continuum point from this work. The solid line is a power law with spectral index $\alpha=-1.7$. The 352~GHz upper limit is from \citet{reu04} and the dashed line shows the Rayleigh-Jeans portion of a modified blackbody spectrum, normalised to the 352~GHz (850~$\mu$m) limit, with T=50~K and $\beta=1.5$. The upper limit at 93~GHz is from this paper using the \co(5-4) flux density averaged over the range 0-300 \kmsec.}
\label{fig4}
\end{figure}

\begin{figure}
\centering
\includegraphics[width=7cm]{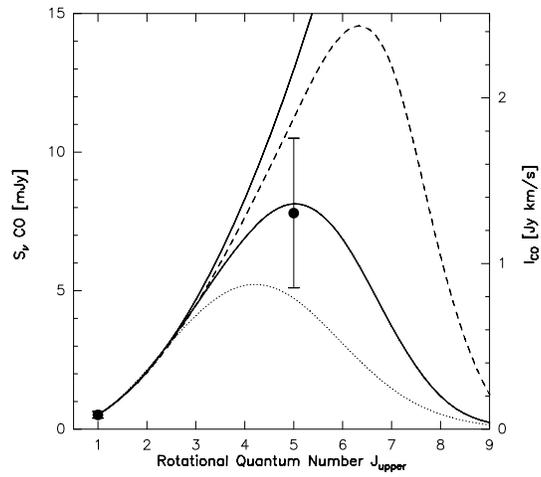}
\caption{\small \co ladder with LVG models overlaid. The best match to the observations is log(n(H$_2))=3.3$, T$_{\rm kin}=50$~K and r$>2.0$~kpc (solid line). Also in agreement with the observations are log(n(H$_2))=4.0$, T$_{\rm kin}=30$~K and r$>2.9$~kpc (dashed line) and log(n(H$_2))=2.7$, T$_{\rm kin}=150$~K and r$>2.2$~kpc. (dotted line). The condition for optically thick, thermally excited gas is the solid, parabolic fit. The error bars are 1$\sigma$ noise estimates and do not include the uncertainty in the absolute flux scale (\S\ref{atcaobservations})}
\label{fig5}
\end{figure}

At $z=5.2$ the universe is only about a billion years old --- time enough, however, to build a supermassive black hole, trigger powerful radio jets and produce a stellar population capable of enriching the interstellar medium with $\sim10^7$ \mdot of \co (assuming an abundance of \co relative to H$_2$ of $Z_{\rm CO}=10^{-4}$). This rapid enrichment could be facilitated in part by the powerful radio jets themselves propagating into overdensities in the primordial gas, and shock-inducing massive star formation \citep{kla04,bic00}.\newline
\citet{ven04} recently postulated that TN~J0924--2201 resides within a protocluster environment after discovering six Lyman-$\alpha$ emitters in a $6\farcm8\times6\farcm8$ field ($2.6\times2.6$~Mpc) around the radio galaxy --- the typical scale size for nearby rich clusters \citep{abe58}. Our \co observations substantiate a five-year old hypothesis that TN~J0924--2201 is a young forming galaxy, potentially the primeval central dominant galaxy of a rich cluster. Thus it seems that cDs may form with their host environments from the earliest times in the Universe.
\acknowledgments
Big thanks go to everyone who made the 12 and 3~mm ATCA upgrades possible. Thanks also to Chris Carilli for the VLA observations which we have incorporated into Figures \ref{fig3} and \ref{fig4} and Christian Henkel who originally wrote the LVG code. IJK acknowledges receipt of an Australian Postgraduate Award and CSIRO Postgraduate Scholarship. The Australia Telescope Compact Array is part of the Australia Telescope which is funded by the Commonwealth of Australia for operation as a National Facility managed by CSIRO.
\clearpage

\clearpage

\clearpage

\clearpage

\clearpage

\clearpage

\begin{deluxetable}{lccc}
\tabletypesize{\scriptsize}
 
\tablecaption{Observed physical parameters of TN~J0924--2201\label{tbl-1}}

\tablewidth{0pt}
\tablehead{
\colhead{} & \colhead{18.6~GHz} & \colhead{\co(1-0)} & \colhead{\co(5-4)}\\
\colhead{} & \colhead{continuum} & \colhead{line} & \colhead{line} }
\startdata

Peak (mJy~beam$^{-1}$) & $0.71\pm0.03$ & $0.52\pm0.12$ & $7.8\pm2.7$\\
Width (\kmsec) & \nodata & $250-400$\tablenotemark{a} & $200-300$ \\
$I_{\textsc{co}}$ (Jy~\kmsec) & \nodata & $0.087\pm0.017$\tablenotemark{a}  & $1.19\pm0.27$\\
$\Delta$RA$_{\rm J2000}$ ($\arcsec$)\tablenotemark{b} & $-0.7\pm0.7$  & $1.5\pm2.1$  & $-1.0\pm1.3$\\
$\Delta$DEC$_{\rm J2000}$($\arcsec$)\tablenotemark{b} & $0.3\pm0.7$ & $4.5\pm1.7$ & $-2.8\pm1.2$ \\

\enddata
\tablecomments{$^a$The weaker feature in the \co(1-0) spectrum has not been included in this measurement because within the 300-400 \kmsec velocity range, there are several other similar features in the image which decrease its significance. If this feature were included the velocity-integrated flux density of the \co(1-0) emission would be $0.12\pm0.03$~Jy~\kmsec.$^b$These positions are offsets from high resolution VLA observations: RA$_{J2000}=09^{\rm h} 24^{\rm m} 19.92^{\rm s}$, DEC$_{J2000}=-22^{\circ} 01\arcmin 41.5\arcsec$ \citep{wvb99}. Positive offsets are North and East of the radio galaxy position.}
\end{deluxetable}

\end{document}